\newcommand{\refcite}[1]{\cite{#1}}

\def\qquad{\hspace{30pt}}
\def\0{\newline}
\def\1{\newline}
\def\2{\par\noindent\newline}

\def\loq{,\kern-0.080em,\kern+0.05em}

\def\bfloq{{\bf,\kern-0.06em,\kern+0.05em}}

\def\footloq{,\kern-0.07em,\kern+0.03em}

\def\quabla{{\raise.7ex\hbox{\boxed{{}}}}}

\documentclass[german,12pt]{article}
\usepackage{epsfig}
\usepackage{url}
 \usepackage[english]{babel}
 \usepackage{latexsym}
 \usepackage{graphicx}
 \usepackage{ifthen}
\title{}
\date{}
\begin{document}

%
%
\begin{center}
\ \\[-1.30cm]
{\LARGE\bf
Reply to
\lq\lq Comment on \lq
Falsification Of\\
The Atmospheric CO$_{\bf 2}$ Greenhouse Effects\\
Within The Frame Of Physics\rq \\
by
Joshua B.\ Halpern,\\
Christopher M.\ Colose,\\
Chris Ho-Stuart,
Joel D.\ Shore,\\
Arthur P.\ Smith,
J\"{o}rg Zimmermann\rq\rq\\[1.00cm]
}
{\large\bf Version 1.00 (December 1, 2010)}\\[1.30cm]
%
%
{\Large\sc Gerhard Gerlich}                                               \\[0.30cm]
{\large\rm Institut f\"ur Mathematische Physik}                           \\[0.30cm]
{\large\rm Technische Universit\"at Carolo-Wilhelmina zu Braunschweig}    \\[0.30cm]
{\large\rm Mendelssohnstra\ss e 3}                                        \\[0.30cm]
{\large\rm D-38106 Braunschweig}                                          \\[0.30cm]
{\large\rm Federal Republic of Germany}                                   \\[0.30cm]
{\large\rm g.gerlich@tu-bs.de}                                            \\[1.30cm]
%
%
{\Large\sc Ralf D. Tscheuschner}                                          \\[0.30cm]
{\large\rm Postfach 60\,27\,62}                                           \\[0.30cm]
{\large\rm D-22237 Hamburg}                                               \\[0.30cm]
{\large\rm Federal Republic of Germany}                                   \\[0.30cm]
{\large\rm ralfd@na-net.ornl.gov}                                         \\[0.30cm]
%
%
\end{center}
%
%
\pagebreak
%
%
\ \\[1cm]
\begin{center}
{\large\bf Abstract}\\
\end{center}
It is shown that the notorious claim by Halpern {\it et al.\/}
recently repeated in their comment that the method, logic, and
conclusions of our \lq\lq Falsification Of The CO$_2$ Greenhouse
Effects Within The Frame Of Physics\rq\rq\ would be in error has no
foundation. Since Halpern {\it et al.\/} communicate our arguments
incorrectly, their comment is scientifically vacuous. In particular,
it is not true that we are \lq\lq trying to apply the Clausius
statement of the Second Law of Thermodynamics to only one side of a
heat transfer process rather than the entire process\rq\rq\ and that
we are \lq\lq systematically ignoring most non-radiative heat flows
applicable to Earth's surface and atmosphere\rq\rq. Rather, our
falsification paper discusses the violation of fundamental physical
and mathematical principles in 14 examples of common
pseudo-derivations of fictitious greenhouse effects that are all
based on simplistic pictures of radiative transfer and their obscure
relation to thermodynamics, {\it including but not limited to\/}
those descriptions (a) that define a \lq\lq Perpetuum Mobile Of The
2nd Kind\rq\rq, (b) that rely on incorrectly calculated averages of
global temperatures, (c) that refer to incorrectly normalized
spectra of electromagnetic radiation. Halpern {\it et al.\/}
completely missed an exceptional chance to formulate a
scientifically well-founded antithesis. They do not even define a
greenhouse effect that they wish to defend. We take the opportunity
to clarify some misunderstandings, which are communicated in the
current discussion on the non-measurable, i.e.\ physically
non-existing influence of the trace gas CO$_2$ on the climates of
the Earth.
\\
\\
Electronic version of an article published as
{\it
International Journal of Modern Physics B,
\mbox{Vol.\ 24},
\mbox{No.\ 10} (2010) 1333--1359\/},
DOI No:\ 10.1142/S0217979210055573,
\copyright\ World Scientific Publishing Company,
\url{http://www.worldscinet.com/ijmpb}.
\pagebreak
%
%
\tableofcontents
\newpage
%
%
\pagestyle{myheadings}
\markboth%
{{\sl
Gerlich and Tscheuschner,
Reply to Comment on \lq\lq Falsification of $\dots$ CO$_2$ $\dots$\rq\rq\ }}%
{{\sl
Gerlich and Tscheuschner,
Reply to Comment on \lq\lq Falsification of $\dots$ CO$_2$ $\dots$\rq\rq\ }}
%
%
\newpage
\section{Introduction}

\subsection{Prologue}
Any statement is subject to a simple question:
\lq\lq Is it true?\rq\rq\

Since Arrhenius (1896)
\cite{Arrhenius1896},
the so called atmospheric greenhouse effect provides a reasoning for
climate change, although his paper (1896)
is {\it arbitrarily wrong\/}.%
\footnote{a formulation used by the theoretical meteorologist Gerhard Kramm
\cite{KrammEMail}.
}
Neither is there any empirical evidence for the existence of an
atmospheric CO$_2$ greenhouse effect, i.e., the influence of the
concentration of the trace gas CO$_2$ on the Earth's climates, nor
there is a definition of an
atmospheric CO$_2$ greenhouse effect in terms of a physical effect
\cite{GT2009,GT2007}.

\subsection{What is a physical effect?}
A physical effect consists of three things:
\begin{quote}
\begin{itemize}
\item[(a)] a reproducible experiment in the lab;
\item[(b)] an interesting or surprising outcome;
\item[(c)] an explanation in terms of a physical theory.
\end{itemize}
\end{quote}
\noindent%
Examples:
\begin{itemize}
\item[(1)]
Hall Effect: The Hall effect
\cite{Hall1879}
is the production of a
potential difference (the Hall voltage) across an electrical
conductor, transverse to an electric current in the conductor and a
magnetic field perpendicular to the current. In the experimental
setups, the strength of the magnetic and electric fields as well as
the mobility of conduction electrons is varied. The effect is
\lq\lq explained\rq\rq\ with the Lorentz force introduced already by
Maxwell. In the meantime the quantum Hall effects have been
discovered whose theoretical \lq\lq explanations\rq\rq\ may be
regarded not completely convincing, although their importance was
recognized by the awarding of several Nobel Prizes in physics (Some
experts say, that the integral quantum Hall effect is less
understood than the fractional quantum Hall
effect)
\cite{Klitzing1980,Tsui1982}.
Clearly, the criteria (a), (b), (c) are fulfilled.
\item[(2)]
The warming process in a car parking in the sun
\cite{GT2009,GT2007}.
Once the interior of the car is heated up the air cooling stops
immediately. This obstruction to air cooling is also at work in case
of fur coats,
blankets, insulating layers etc.%
\footnote{%
\lq\lq Every climatologist should ask himself/herself: Why is the
sparrow on a cold morning not freezing to death?\rq\rq\ (after
Wolfgang Th\"{u}ne)}
This can be explained without any physical skills
as it was already known by the Neanderthalian,
which was formulated firstly by the popular German meteorologist Wolfgang Th\"{u}ne.
Evidently, a class of certain verifiable processes %
reproducible by measurements (a) are given. %
However, there is no non-trivial physical explanation, %
cf.\ (b) and (c). Therefore, it is justified to
christen this (non-physical) effect
\lq\lq Neanderthalian effect\rq\rq.
\item[(3)]
Sometimes one describes by the natural greenhouse effect the
circumstances, that without the trace gases (carbon dioxide etc.)
the global average temperatures of the atmosphere near ground would
have minus 18 degrees Celsius. Evidently, property (a) is not
fulfilled, since there are no reproducible and comparable
measurements. Therefore, the so called natural greenhouse effect is
not a physical effect. It was called a \lq\lq meteorological
effect\rq\rq%
\footnote{Not all meteorologists would agree.}
by the first author in his Leipzig talk
\cite{GerlichLeipzig}.
\end{itemize}
Hence there are no greenhouse effects in
physics
\cite{GT2009,GT2007}.
Beyond this simple observation, there
is a falsification of the atmospheric CO$_2$ greenhouse effects in
the two senses of this homonymous word:
\begin{itemize}
\item it is a fake within the framework of so-called climate science,
\item it is falsified in a Popperian sense within the frame of
     physics.
\end{itemize}
This is one main result of our paper
\cite{GT2009,GT2007}.
The other
main result is that the concentration of carbon dioxide has no
measurable influence on the temperature field of the atmosphere of
the Earth
\cite{GT2009,GT2007}.

\subsection{%
The comment by Halpern}
The results of our paper are not the results
of (so-called) climate science or chemistry, but of theoretical and
applied physics. Therefore, the submission of our article to an
applied physics journal did make sense. In our honest opinion this
is not true for the recent comment by the chemist Halpern and his
co-authors
\cite{Halpern2010}.%
\footnote{See also
Refs.~\refcite{rabett.blogspot.com} and \refcite{NYT}.}

 We do not agree at all that our
\begin{quote}
\begin{itemize}
\item \lq\lq methods, logic, and conclusions are in error.\rq\rq\ \end{itemize}
\end{quote}
To our surprise Halpern {\it et al.\/}\ {\bf did not even define a
greenhouse effect}, such that their work is scientifically
worthless, since, without a sharp definition of the concept in
question, any scientific comment or any scientific refutation is
impossible. The two core statements of Halpern {\it et al.\/}
\begin{quote}
\begin{enumerate}
\item[($\textrm{H1}$)]
that we are \lq\lq trying to apply the Clausius statement of the
Second Law of Thermodynamics to only one side of a heat transfer
process rather than the entire process\rq\rq\ and we are \lq\lq
systematically ignoring most non-radiative heat flows applicable to
Earth's surface and atmosphere\rq\rq;
\item[($\textrm{H2}$)]
that we claim that \lq\lq radiative heat transfer from a colder
atmosphere to a warmer surface is forbidden, ignoring the larger
transfer in the other direction which makes the complete process
allowed by ignoring heat capacity and non-radiative heat flows they
claim that radiative balance requires that the surface cool by 100 K
or more at night\rq\rq;
\end{enumerate}
\end{quote}
are incorrect.%
\footnote{%
In order to verify the curious reader is recommended to activate the
\lq\lq search and find\rq\rq\ option of his favorite text viewing software.%
} Rather, we show
\cite{GT2009,GT2007}
\begin{quote}
\begin{enumerate}
\item[($\overline{\textrm{H1}}$)]
that some pseudo-explanations of a fictitious atmospheric natural
greenhouse effect or atmospheric CO$_2$ greenhouse effect describe a
Perpetuum Mobile of the Second Kind and, thus, violate the Clausius
law;
\item[($\overline{\textrm{H2}}$)]
that many discussions which speculate on an influence of the
concentration of the trace gas CO$_2$ on the climates only rely on a
simplistic discussion of radiative transfer, while ignoring heat
conductivity, convection, friction, interface physics.
\end{enumerate}
\end{quote}
In other words, we analyze the rationale and the inner contradiction
of derivations of the atmospheric greenhouse effects communicated in
the standard climate literature from the viewpoint of a physicist.
In part, we are arguing within the context of the standard
assumptions put forward by mainstream global climatologists. Nowhere
we offer our own model, and we never will do.

\subsection{This paper}
We have made time and have tried to trace back the origins of the
objections raised against our paper. The rest of our response should
clarify these misunderstandings. However, we cannot repeat our
previous work here, to which the reader is
referred
\cite{GT2009,GT2007}.

One should keep in mind that we are theoretical physicists with
experimental experience and, additionally, a lot of experience in
numerical computing.
Joshua Halpern%
\footnote{%
{\it Note added:\/}
Dr.\ Joshua B.\ Halpern
({\it alias\/} blogger Eli Rabett),
Professor of Chemistry at Howard University, Washington DC,
is a physicist by education,
{\it cf.\/}
\url{http://www.coas.howard.edu/chem/jhalpern/}.%
}
and J\"{o}rg Zimmermann,
for example, are chemists.

We are not willing to discuss whether they
can be considered as laymen in physics, in particular laymen in
thermodynamics.%
\footnote{However, we must think so.}
\newpage
\section{%
Some General Remarks on Statements Appearing
in the Comment by Halpern et al.}
\subsection{%
Basic facts%
}
The title of our paper reads: \lq\lq Falsification Of The
Atmospheric CO$_2$ Greenhouse Effects Within The Frame Of
Physics\rq\rq. Until now, there are no papers that refute our work.
The only attempt to try this so far is due to Arthur P. Smith
(2008)
\cite{Smith2008}.
However, Kramm, Dlugi, and Zelger (2009)
showed that his entire paper is wrong
\cite{Kramm2009}.
Smith used
inappropriate and inconsistent formulations in averaging various
quantities over the entire surface of the Earth considered as a
sphere. Using two instances of averaging procedures as customarily
applied in studies on turbulence, Kramm, Dlugi, and Zelger show that
Smith's formulations are highly awkward. In their work, Kramm,
Dlugi, and Zelger scrutinize and evaluate Smith's discussion of the
infrared absorption in the atmosphere. They show that his attempt to
refute our criticism is rather fruitless. The same holds true for
the comment of Halpern {\it et al.\/}
\cite{Halpern2010}.
Qualified readers as well as laymen can verify the invalidity of
many of the claims communicated by Halpern {\it et al.\/} simply by
using the \lq\lq search and find\rq\rq\ option of their document
reader. Therefore, in what follows, we restrict ourselves to list
some general remarks on the physics related to the statements
appearing in the paper by Halpern {\it et al.\/}\,. Some more important
topics are treated more comprehensively below in separate sections,
namely
\begin{itemize}
\item %
the Clausius law and the related errors of Rahmstorf, Hoffmann,
Halpern {\it et al.\/}, Ozawa {\it et al.\/} in
Section~\ref{Clausius};
\item %
the radiation spectra and the related errors of Bakan and Raschke in
Section~\ref{BR};
\item %
the adiabatic lapse rate (barometric formula) and the related claims
of Rahmstorf and Schellnhuber that Venus suffers from an atmospheric
CO$_2$ greenhouse effect in Section~\ref{lapse}.
\end{itemize}
We find that these points are very important, since, once again,
they refute the greenhouse myth underlying the mainstream view of
the influence of CO$_2$ on the climates. We as physicists emphasize:
The \lq\lq mainstream\rq\rq\ view is clearly
wrong
\cite{GT2009,GT2007,Kramm2009}.

\subsection{%
Some remarks on Section~1 (Introduction)%
}
Let us start with Halpern's favorite object of lust
\cite{Halpern2010,rabett.blogspot.com}.
In our falsification paper we
criticize the suggestive abuse of a graphical language by global
climatologists
\cite{GT2009,GT2007}.
This is very important in the
case of radiation balance diagrams, since one must never confuse
classical radiation intensities, energy flows, and heat flows. For
instance, Rahmstorf himself charismatically confuses energy and
heat
\cite{RahmstorfHomepage}:
\begin{quote}
Manche "Skeptiker" behaupten, der Treibhauseffekt k\"{o}nne gar nicht
funktio\-nieren, da (nach dem 2.\ Hauptsatz der Thermodynamik) keine
Strahlungsenergie von k\"{a}lteren K\"{o}rpern (der Atmosph\"{a}re) zu w\"{a}rmeren
K\"{o}rpern (der Oberfl\"{a}che) \"{u}bertragen werden k\"{o}nne. Doch der 2.
Hauptsatz ist durch den Treibhauseffekt nat\"{u}rlich nicht verletzt, da
bei dem Strahlungsaustausch in beide Richtungen netto die Energie
von warm nach kalt flie{\ss}t.
\end{quote}
This may be translated to
\begin{quote}
Some \lq\lq sceptics\rq\rq\ state that the greenhouse effect cannot
work since (according to the second law of thermodynamics) no
radiative energy can be transferred from a colder body (the
atmosphere) to a warmer one (the surface). However, the second law
is not violated by the greenhouse effect, of course, since, during
the radiative exchange, in both directions the net energy flows from
the warmth to the cold.
\end{quote}
This is not a quotation out of context, it is plainly wrong, since
it confuses \lq\lq radiative energy\rq\rq\ and \lq\lq heat\rq\rq\ in
such a way that the brainwashed reader is losing all orientations.

Normally, arrows indicate that the relevant physical quantities are
flows (vector fields) that can be superposed due to their inherent
linear structure. The intensities of classical radiation theory are
not flows. In addition, energy flows, in general, and heat flows, in
particular, have to be strictly distinguished in the context of a
thermodynamical analysis. Moreover, in the movie by Al
Gore
\cite{AlGore1,AlGore2}
there are diagrams, reminiscent of wave
reflection. This is nonsense too. In mainstream graphical
representation all this is mixed together, and, in addition, reduced
to a one-dimensional view far from any reality
\cite{GT2009,GT2007}.

Let us now discuss the diagrams in Fig.\ 23, p.\ 322 in our
paper
\cite{GT2009,GT2007}.
We do not discuss only one option for the
interpretation of such diagrams, as is suggested by the objection of
Georg Hoffmann
\cite{HoffmannBlog}
and Joshua Halpern {\it et al.\/}
\cite{Halpern2010,rabett.blogspot.com}.
Rather, we discuss four
possible interpretations, that were introduced for
paedagogical/didactical reasons to emphasize that any diagrammatic
language in science has to have a well-defined syntax and
semantics
\cite{GT2009,GT2007}.
Once again: The mentioned examples,
e.g.\ \lq\lq Feynman diagrams\rq\rq, \lq\lq SysML\rq\rq, should only
remind the reader of the important fact that a graphical language in
science should always have a well-defined syntax and semantics. This
was the problem with Pauli's negative opinion on the graphical
language introduced by St\"{u}ckelberg. This language was later refined
by Feynman, and the end of the story is well-known. Pauli called
St\"{u}ckelberg's ideas \lq\lq Stimmungsmalerei\rq\rq\ (Painting of
moods)
\cite{Heinz}.
That is what it exactly is - in the case of
radiation balance diagrams! In the case of St\"{u}ckelberg-Feynman
diagrams, these graphic representations could be integrated into the
rigorous formalism of Green functions,
pioneered by Julian Schwinger.%
\footnote{%
Many years ago, one of us (R.D.T.) had the opportunity, to discuss
the issue of graphical representation with Professor Schwinger and
his amusing rivalry between Feynman and him.}
The introduction and application of graphical languages is an
interesting topic in informatics and related to mathematical
problems such as graph theory, knot theory and so on. On the other
hand, there are many other fields, where a poorly defined graphical
language is used, e.g.\ in business related topics, and, of course,
in all kinds of brainwashing.

Radiation balance diagrams, however, are really useless. They never
occurred in the talks of the first author (G.G.), who takes the
opportunity and freedom to add a much simpler argument here:
Especially in the diagram depicted in the paper one cannot find one
single ratio (in percent) that is a ratio of measured numbers!

Furthermore, we remark in the context of Section~1 of
Ref.~\refcite{Halpern2010}:
\begin{quote}
\begin{itemize}
\item[(a)]
Halpern {\it et al.\/} confuse Global Climate Models (which they
abbreviate as GCM) and General Circulation Models (GCM), i.e.\
coupled atmosphere-ocean models. By global climatologists, the
latter are considered as the \lq\lq key components\rq\rq\ of the
former, whatever this means, but not identified with each other.
\item[(b)]
It is impossible to calculate temperature fields of the Earth's
atmosphere by using radiative transfer equations regardless of an
introduction of CO$_2$ concentration or molecule spectra, let alone
\lq\lq line-by-line\rq\rq\ and/or \lq\lq state-of-the-art\rq\rq\
calculations.
\item[(c)]
The critique that we rely on unrefereed sources is distorting the
facts; most of our citations are peer-reviewed or from classical
textbooks. And if not, then it will have its own reason.
\item[(d)]
Halpern {\it et al.\/} intentionally misunderstand and exaggerate
side remarks (as  shown above) in order to discredit us.
\item[(e)]
It is true that the heat conductivity of a gas is relatively small.
However, it is still finite. Heat conductivity plays an important
role at the interface between ground and atmosphere and, of course,
serves as a germ for heat transfer by convection. The latter
surmounts ordinary (static) heat conductivity typically by four
orders of magnitude.%
\footnote{%
That is why our soup becomes cold when the door is left open. The
same effect happens during bake-out of donuts in sizzling oil: The
cook ends up in a screaming frenzy.}
\item[(f)]
Contrary to what Halpern {\it et al.\/} state, %
we emphasize the importance
of the non-radiative forms of heat transfer including convection and
latent heat, e.g.\ already in Section~1.2.\ of our
paper
\cite{GT2009,GT2007}.
\end{itemize}
\end{quote}

\subsection{%
Some remarks on Section~2
(The greenhouse effect and the second law of thermodynamics)%
} Some common misunderstandings related to the Second Law of
Thermodynamics are discussed below in a Section~\ref{Clausius}.
At this point we emphasize
\begin{itemize}
\item[(a)]
We never claimed - allegedly with reference to Clausius - that a
colder body does not send radiation to a warmer one. Rather, we cite
a paper, in which Clausius treats the radiative
exchange
\cite{Clausius1887a,Clausius1887b}.
The correct question is, whether
the colder body that radiates less intensively than the warmer body
warms up the warmer one. The answer is: It does not.
\item[(b)]
Speculations that consider the conjectured atmospheric CO$_2$
greenhouse effect as an \lq\lq obstruction to cooling\rq\rq\
disregard the fact that in a volume the radiative contributions are
already included in the measurable thermodynamical properties, in
particular, transport coefficients. These will show no measurable
variations if one doubles the CO$_2$ concentration. Furthermore, the
\lq\lq obstruction models\rq\rq\ often neglect the fact that \lq\lq
radiative balance\rq\rq\ is introduced as a preposition of the
standard analysis.
\item[(c)]
We repeat a statement from above: It is true that the heat
conductivity of a gas is relatively small. However, it is still
finite. Heat conductivity plays an important role at the interface
between ground and atmosphere and, of course, serves as a germ for
heat transfer by convection. The latter surmounts ordinary (static)
heat conductivity typically by four
orders of magnitude.%
\item[(d)]
Of course, heat conductivity of the ground is non-negligible.
Halpern {\it et al.\/} should read our paper more carefully. By the
way, the pot-on-the-stove example, only shows that infrared
absorption and heat conductivity are not related to each other.
\item[(e)]
The Stefan-Boltzmann $T^4$-law does only apply to an idealized black
body, that is a cavity with a hole placed in a heat bath of constant
temperature $T$. Global climatologists use crude approximations,
from which they compute tiny variations of measurable quantities
unscrupulously. This is inadmissible. One example is the conjectured
atmospheric CO$_2$ greenhouse effect. Even if their theory were
correct the error bars would render their predictions useless, since
being gigantic.
\item[(f)]
A so-called grey body obeying a modified Stefan-Boltzmann $T^4$ law
(i.e. a Stefan-Boltzmann law multiplied by a factor) is a
phenomenological construct whose physical realization does not
exist.
\item[(g)]
The Earth is a multi-colored object characterized by an inhomogenous
color distribution, not a black or grey body, which cannot be
altered by Arthur B.\ Smith, who essentially has plagiarized our
inequality
\cite{Smith2008}
and did not refute anything of our
work
\cite{GT2009,GT2007,Kramm2009}.
By using the Stefan-Boltzmann
law one always computes radiations that are far too
large
\cite{GT2009,GT2007,Kramm2009}.
\item[(h)]
Gaseous layers never obey the Stefan-Boltzmann $T^4$ law. All these
calculations (e.g.\ the shell layer calculations performed in detail
by Halpern {\it et al.\/}) are fundamentally wrong and prove
nothing. The corresponding four pages of the comment by Halpern {\it
et al.\/} are obsolete.
\item[(i)]
If one introduces discretizations (lattice cells, finite number of
layers) one must always discuss either the continuum limit or the
artifacts generated by the discretization thoroughly. The \lq\lq
philosophy\rq\rq\ communicated by the numerical mathematician and
global
climatologist von Storch%
\footnote{%
It should be noted that von Storch was one of the first global
climatologist who refuted the \lq\lq Hockey Stick\rq\rq\ by Michael
Mann {\it et al.\/} However, as his textbook shows, he still accepts
the atmospheric CO$_2$ greenhouse hypotheses.}
\lq\lq The discretization is the model\rq\rq\
\cite{vStorch}
is not
only simplistic but fundamentally unphysical.%
\footnote{%
A nice example is the comparison of the discrete and continuous
versions of the logistic equations (Verhulst, Feigenbaum).}
\end{itemize}

\subsection{%
Some remarks on Section~3
(A rotating planet etc.)%
} This section presents nothing new. The reader is referred to our
paper and the work of Kramm, Dlugi, and Zelger
\cite{Kramm2009}.
We emphasize:
\begin{itemize}
\item[(a)]
Repeating our statement from above, it is impossible to calculate
temperature fields of the Earth's atmosphere by using radiative
transfer equations regardless of an introduction of CO$_2$
concentration or molecule spectra.
\item[(b)]
The radiative transfer equations do not yield the portion of
radiation energy that is transformed into heat. This can be easily
seen by observing that the direction of the gradient of the
temperature determines whether the lines of the spectrum are present
as absorption lines (Fraunhofer lines) or emission lines. In case of
the so called scattering atmosphere after
Chandrasekhar
\cite{Chandrasekhar}
no portion of the radiation energy
is thermalized at all.
\item[(c)]
It is impossible to measure the temperature fields of the Earth's
atmosphere or any warming effect in spectroscopic experiments.
Halpern {\it et al.\/} do not prove their assertion stated in
Section~3.4 of
Ref.~\refcite{Halpern2010}
that the \lq\lq downward
emission\rq\rq\ term is \lq\lq by a factor of roughly two\rq\rq\
larger than the incident solar radiation. The origin of the Planck
curves in the Fig.\ 7 of
Ref.~\refcite{Halpern2010}
is rather
obscure. Taken seriously, it would mean that the detectors are
gauged with help of idealized black body measurements.
\item[(d)]
Again: %
We never claimed - allegedly with reference to Clausius -
that a colder body does not send radiation to a warmer one. Rather,
we cite a paper, in which Clausius treats the radiative
exchange
\cite{Clausius1887a,Clausius1887b}.
The correct question is, whether
the colder body that radiates less intensively than the warmer body
warms up the warmer one. The answer is: It does not.
\end{itemize}

\subsection{%
Some remarks on Section~4
(Climate models)%
} %
Halpern {\it et al.\/} correctly recognize that, in our opinion,
global climate models and the study of scenarios, do not belong to
the realm of science
\cite{GT2009,GT2007}.
To put it bluntly, they
are science fiction. Their review of climate models reminisces what
can be read in the mainstream literature. and presents nothing new.
Halpern {\it et al.\/} find it inappropriate that we discuss some
fundamentals of the philosophy of science in the context of our
paper. However, it is important, to remember that science is a
method to test hypotheses. We would be glad if Halpern {\it et
al.\/} conclusively explained why the predictions of different
climate models differ fundamentally and miss the reality completely.
From a physicist's point of view, we should emphasize:
\begin{itemize}
\item[(a)]
Repeating our statement from above, Halpern {\it et al.\/} confuse
Global Climate Models (which they abbreviate as GCM) and General
Circulation Models (GCM), i.e.\ coupled atmosphere-ocean models. By
global climatologists, the latter are considered as the \lq\lq key
components\rq\rq\ of the former, whatever this means, but not
identified with each other.
\item[(b)]
The Navier-Stokes equation has a friction term. Without this term,
this equation becomes the Euler equation. With a friction term, the
velocity field obeys a different boundary condition than without.%
\footnote{cf. Ludwig Prandtl's interface layer}
In case of a friction term one needs the second derivatives of the
velocity fields that cannot be approximated with help of the wide
mesh lattices of the climate models. The same fact holds for the
heat conduction equation. In our paper, we emphasize that even the
simplest form of time evolution equations for the Earth (atmosphere
and oceans) cannot be treated numerically near reality.
\end{itemize}
Thus, global climate models are nothing but a very expensive form of
computer game entertainment.
\subsection{%
Some remarks on Section~5
(Systematic problems: Definition of the
greenhouse effect, assertions,
theoretical arguments)%
}
In our falsification paper
\cite{GT2009,GT2007}
we discuss
different versions of the greenhouse effect which, in part,
contradict to each other.
\begin{itemize}
\item[(a)]
As already emphasized, Halpern {\it et al.\/}
do not choose from the existing versions
of the greenhouse effect nor define their own one which they prefer to defend.
Thus the comment of Halpern {\it et al.\/}
is scientifically worthless.
\item[(b)]
Again: %
It is impossible to calculate temperature fields of the Earth's
atmosphere by using radiative transfer equations regardless of an
introduction of CO$_2$ concentration or molecule spectra.
\item[(c)]
Again: %
The radiative transfer equations do not yield the portion of
radiation energy that is transformed into heat. This can be easily
seen by observing that the direction of the gradient of the
temperature determines whether the lines of the spectrum are present
as absorption lines (Fraunhofer lines) or emission lines. In case of
the so called scattering atmosphere after
Chandrasekhar
\cite{Chandrasekhar}
no portion of the radiation energy
is thermalized at all.
\item[(d)]
Contrary to the claims of Halpern {\it et al.\/},
the system of
equations discussed in the Section~4 of our paper, \lq\lq Physical
Foundations of Climate Science\rq\rq\
\cite{GT2009,GT2007},
is entirely relevant as it includes the oceans, the stratosphere, the
electrodynamics, and so on. Halpern {\it et al.\/}
try to channelize
the discussion by arbitrarily labeling issues as relevant or
irrelevant. Do the authors of the comment have a reason
which makes them sure to know enough?
\item[(e)]
Contrary to the claims of Halpern {\it et al.\/},
the equations of
magnetohydrodynamics, and in particular, electrodynamics belong to
the physical basis of the atmospheric problem. They are relevant to
the description of clouds, thunder and lightning, electromagnetic
radiation, and in particular, dielectric properties of the
components of the atmosphere.
\item[(f)]
The beloved CO$_2$ is a dielectricum. Not only physicists like Georg
Hoffmann should know the consequences with regard to the Maxwell
equations and Beer's law
\cite{Beer},
namely one has to distinguish
between scattering and (true) absorption. Prominent astrophysicists
as Chandrasekhar (Chicago) and Uns\"{o}ld (Kiel) have elaborated on this
difference. Global climatologists should become more familiar with
the work of these giants
\cite{Chandrasekhar,Unsoeld}.
\end{itemize}
\newpage
\section{Halpern et al.\ versus Clausius}
\label{Clausius}

\subsection{Objections adapted from Georg Hoffmann}
In the acknowledgement of their paper Halpern {\it et al.\/}\ thank
Georg Hoffmann (among others) for his suggestions.
Georg Hoffmann argues we would state that \lq\lq there is
no greenhouse effect, that this effect contradicts to the second law
of thermodynamics and climate modelers do not know anything about
physics\rq\rq\
\cite{HoffmannBlog}.
The quotes indicate that this quotation is supposed
to be {\it verbatim.\/} However, one cannot find this in the text of
the falsification paper
\cite{GT2009,GT2007}.
In order to verify, one
only needs to activate \lq\lq search and find\rq\rq\ inputting the
corresponding search terms.

Naturally, from our own experience we know - and we often point this
out in discussions - that individuals, who - escaped from the
science department - flew to and finally got lost in the domains of
global climatology often suffer from a barely modest infection by
mathematics and physics. For instance, Georg Hoffmann apparently
does not know how to apply the second law of thermodynamics. The
second law is not a real process that is forbidden, its description,
however, is!

\subsection{Descriptions that contradict
the second law of thermodynamics}
In our paper, we explicitly isolate those descriptions that
contradict the second law of thermodynamics. Of course, there are
some descriptions that do {\bf not} contradict to the second law.
For instance, it suffices to remove only a single sentence in the
proposal of Dipl.-Phys.\ Professor Dr.\ Peter Stichel
\cite{Stichel}:
\begin{quote}
\lq\lq Now it is generally accepted textbook knowledge that the
long-wave infrared radiation, emitted by the warmed up surface of
the Earth, is partially absorbed and re-emitted by CO$_2$ and other
trace gases in the atmosphere. This effect leads to a warming of the
lower atmosphere and, for reasons of the total radiation budget, to
a cooling of the stratosphere at the same time.\rq\rq\ \end{quote}
However, in its original form, it describes a Perpetuum Mobile of
the Second Kind. We repeat our statement from above:
\begin{quote}
\begin{itemize}
\item
Once again, we never claimed - allegedly with reference to Clausius - that a
colder body does not send radiation to a warmer one. Rather, we cite
a paper, in which Clausius treats the radiative
exchange
\cite{Clausius1887a,Clausius1887b}.
The correct question is, whether
the colder body that radiates less intensively than the warmer body
warms up the warmer one. The answer is: It does not.
\end{itemize}
\end{quote}
Thus the critique of Halpern {\it et al.\/} does not apply.

\subsection{Four examples of objections against our discussion
of the fictitious greenhouse effects and the second law}
\subsubsection{The argument by Halpern et al.}
According to Halpern {\it et al.\/}
\cite{Halpern2010}
\begin{quote}
It is not admissible \lq\lq to apply the Clausius statement of the
Second Law of Thermodynamics to only one side of a heat transfer
process rather than the entire process\rq\rq\ \end{quote}
\subsubsection{An argument by Rahmstorf}
According to Rahmstorf
\cite{RahmstorfHomepage}
\begin{quote}
\lq\lq the second law is not violated by the greenhouse effect, of
course, since, during the radiative exchange, in both directions the
net energy flows from the warmth to the cold.\rq\rq\ \end{quote}

\subsubsection{An argument by Hoffmann}
According to Georg Hoffmann
\cite{HoffmannEMail}
\begin{quote}
\lq\lq 2nd law is always a statement on net heat flows. To consider
only one part of the exchange is incorrect.\rq\rq\ \end{quote}
\subsubsection{An argument by Ozawa et al.}
In their paper \lq\lq The Second Law Of Thermodynamics And The
Global Climate System: A Review Of The Maximum Entropy Production
Principle\rq\rq\ Ozawa {\it et al.\/} write
\cite{Ozawa}:
\begin{quote}
\lq\lq This is not a violation of the second law of thermodynamics
since the entropy increase in the surrounding system is
larger.\rq\rq\ \end{quote}

\subsection{The work of Ozawa et al.}
Comparing these four
examples
\cite{Halpern2010,HoffmannBlog,RahmstorfHomepage,Ozawa},
one observes that there is a confusion about the division of the world
into a system and an environment, and how to handle the basic
concepts. In particular, a discussion
of the errors of Ozawa {\it et al.\/}
clarifies these widespread misunderstandings.

Firstly, one observes, that what Ozawa {\it et al.\/}
write on the first page about Carnot is not true.
Apparently, the authors did not
read the original work of Carnot
\cite{Carnot}:
He postulated the
conservation of heat. According to Carnot, work was produced when
heat drops from the higher temperature level to the lower level.  He
did not transform heat into work. The so-called \lq\lq
principle\rq\rq\ of maximum entropy production is not the second
fundamental law of thermodynamics. We give the correct
formulation
\cite{GT2009,GT2007},
which was given by the inventor of
entropy, Prof.\ R.\ Clausius
\cite{Clausius1887a,Clausius1887b},
who gave the mathematical formulations of the first and second
fundamental law of thermodynamics.

However, to explain the additional mistakes in this paper is much
more difficult, because even in very famous textbooks of Theoretical
Physics the formulation of the second fundamental law of
thermodynamics often is wrong, for instance, in the book \lq\lq
Statistical Physics\rq\rq\ of Landau and Lifshitz
\cite{LandauLifshitz}.%
\footnote{%
One of us G.G.\ intents to write a collection of brief textbooks in
theoretical physics with the main title \lq\lq The Mathematical
Principles and Methods of Physics\rq\rq\ based on his lectures given
at Braunschweig Technical University. At present, only the lecture
notes can be downloaded and cited
\cite{GerlichLectureNotes}.}

A short additional remark concerning Carnot: We are sorry to say,
that everything that Ozawa {\it et al.\/}
write  about Carnot (in
the whole paper) is incorrect.  For instance in Section~2 of
Ref.~\refcite{Ozawa}
the authors write something about Carnot, which
one cannot find in Carnot's treatise: Carnot did not study the earth
as a heat engine, but he studied the theoretical description of
steam engines (cf.\ pp.\ 9-14 of
Ref.~\refcite{GerlichThermodynamik}).
Furthermore, one cannot find - in the entire paper - a correct
formulation of the second fundamental law of thermodynamics and a
formulation of the maximum entropy production principle
\cite{Ozawa}.

All three statements in
Ref.~\refcite{Ozawa},
Section~8
\begin{quote}
\begin{enumerate}
\item
\lq\lq The second law (the law of entropy increase) is valid for a
whole (isolated) system.\rq\rq\ \item \lq\lq When we sum up all the
changes of interacting subsystems, the total change must be
nonnegative.\rq\rq\ \item \lq\lq This is the statement of the second
law of thermodynamics.\rq\rq\ \end{enumerate}
\end{quote}
are wrong. One can find the correct formulation of the second
fundamental law in Section~3.9.1.\ of
Ref.~\refcite{GT2009,GT2007}.
One can formulate with both fundamental laws of thermodynamics
inequalities which are similar to the given statements, but with
more assumptions and constraints, not about the entropy, but about
the sum of entropies. One can find the correct formulations in
Ref.~\refcite{GerlichThermodynamik},
p.50-52.

The definition of the entropy (of a system) is incomplete, thus
wrong. In the formula (1) of
Ref.~\refcite{Ozawa},
Section~8
\begin{equation}
dS = \frac{ dQ }{ T }
\end{equation}
the important point is that $dQ$ is the {\it reversible\/}
differential form of the heat exchange and $T$ is the absolute
temperature, say
\begin{equation}
dS = \frac{ dQ^{rev} }{ T }
\end{equation}
The entropy change is calculated for one system, which has only one
temperature. Within the context of classical thermodynamics one only
has one value for the change of the temperature, of the internal
energy, of the free energy, volume, density, entropy and so forth.
When one would like to use functions of space and time one has to go
over to the field description of irreversible thermodynamics for
instance hydromagnetics.

The entropy production equation (for the sum of entropies) in our
preprint is Equation (143) in
Refs.~\refcite{GT2009,GT2007}.
One has
to integrate this equation over a finite volume. Then one has to
take into account boundary conditions for the surface integrals.
With the first fundamental law of thermodynamics one gets the
inequality (in linear approximation)
\begin{equation}
dQ^{rev} \ge dQ
\end{equation}
which follows from
\begin{equation}
dW^{rev} \ge dW
\end{equation}
where $dW$ is the differential form of the outer work (e.g.\
$dW^{rev} = p\,dV$), cf.\
Ref.~\refcite{GerlichThermodynamik},
p.\ 21.
These inequalities give the inequalities of the sums of
entropies
Ref.~\refcite{GerlichThermodynamik},
pp.\ 49--53. There are
very similar looking inequalities
Ref.~\refcite{GerlichThermodynamik},
p.\ 49. The first one is a
statement about the sum of entropies of two systems:
\begin{equation}
\Delta S_1 + \Delta S_2 > 0
\end{equation}
We quote from
Ref.~\refcite{GerlichThermodynamik},
p.\ 49:
\begin{quote}
Diese Summenentropie des aus zwei Systemen zusammengesetzten
Systems, das insgesamt keine W\"{a}rme und Arbeit austauscht, kann nur
zunehmen, bis sie maximal ist und die Teilsysteme die gleiche
Temperatur haben.
\end{quote}
which may be translated to
\begin{quote}
The sum entropy of the system which is composed of two sub-systems
and which does not exchange neither net heat or net work, can only
increase so long, until it reaches a maximum and both sub-systems
have the same temperature.
\end{quote}
Without additional assumptions one cannot prove these inequalities
for more than two systems
Ref.~\refcite{GerlichThermodynamik},
p.\ 51.
Then, there is another similar looking inequality
Ref.~\refcite{GerlichThermodynamik},
p.\ 49
\begin{equation}
\frac{Q_1}{T_1} + \frac{Q_2}{T_2} \le 0
\end{equation}
In our case, we have only one system, which exchanges heat at two
temperatures and is back in the initial state. One gets the equality
when both heat exchanges are reversible. One could not find in
Ref.~\refcite{Ozawa}
a precise definition of the \lq\lq maximum
entropy production principle\rq\rq. Usually one has constraints and
boundary conditions, when one formulates a variational principle. It
is necessary to give the region of the possible states, which are
allowed. Otherwise it is undefined.
\newpage
\section{The work of Bakan and Raschke}
\label{BR}
\subsection{A review of mistakes}
Some time ago, the first author of the falsification
paper
\cite{GT2009,GT2007}
was concerned with the work of Bakan and
Raschke (on the so-called natural greenhouse effect)
\cite{BR},
because on its first two pages so many mistakes can be found that it
never should pass the referee process. At that time, the climate
hysterics sharply attacked
\cite{BeckToTheFuture,WeirdestMillenium,CurveManipulation}
CO$_2$-history expert Ernst-Georg Beck
\cite{Beck2006},
since he
inadvertently omitted the units of the ordinate that probably got
lost during layout.
This has been inadequately characterized as a scientific fraud.%
\footnote{%
Recently, these false accusations have been repeated
\cite{NewRahmstorfBlog}}.
\\
\\
But what is then the paper by
Bakan and Raschke
\cite{BR}?
\\
\\
The first author informed Dr.\ Eberhard Raschke that in his joint
paper co-authored with Dr.\ Stefan Bakan
\cite{BR}
also the ordinate
unit is missing, a very crucial point with respect to the
quantification of their greenhouse hypothesis. Other serious errors
in
Ref.~\refcite{BR}
are:
\begin{itemize}
\item[(1)]
The year of Fourier's paper is 1824 rather than 1827.
This is important since this error winds its way through
the literature. It leaves a trace of experts behind it
who never read Fourier's paper that does not describe
the greenhouse effect contrary to what is communicated.
\item[(2)]
The contents of the papers by Fourier and Arrhenius is summarized
incorrectly. On page 320 of the IJMPB version of the falsification
paper
\cite{GT2009}
one will find in the facsimile of the text by
Arrhenius that the temperature rise for a doubling of the carbon
dioxide concentration is 1.6 degrees Celsius rather than 6 degrees
Celsius.
\item[(3)]
In the figure caption of the diagram Fig.\ 2-1, p.\ 2, the
temperature for the ground is given by 250 K, whereas in the
accompanying text it reads 255 K.
\item[(4)]
Equal areas do correspond to equal intensities rather than energy
quantities.
\item[(5)]
In the diagrams of Fig.\ 2-1, p.\ 2, the black body Planck curves
are never drawn, as stated, since the curves associated with the
lower temperatures always lie below the curves associated with the
higher temperatures. Without specifying a norm the adjective \lq\lq
normalized\rq\rq\ has no meaning. The Planck function for the
surface of the sun has to be reduced by a factor given by the ratio
earth orbit per sun radius (i.e.\ 46225 = 215~squared) if one
prefers to compare them with the radiation of the ground. Even in
this representation these curves do not look equal. Choosing the
temperatures 5780 K and 290 K, as done in the falsification
paper
\cite{GT2009,GT2007},
one has to scale down the intensity of
the solar radiation by 3.5 (cp.\ Fig.\ 13, p.\ 295, l.h.s.). In case
of the temperatures chosen by Bakan and Raschke themselves this
factor rises to 7.1, which one can easily prove with a standard
computer program. An obfuscation of this necessary rescaling (3.5
resp.\ 7.1) is a suggestive deception because the maximum radiation
of the ground is much less than the incoming solar radiation.
\item[(6)]
Hence, it will remain Georg Hoffmann's personal secret, to which
extent this diagram \lq\lq visualizes beautifully\rq\rq\ Wien's
displacement law. To put it bluntly, this diagram does not visualize
Wien's displacement law at all! This can be verified with rather
elementary mathematics skills. Namely, it is not simply the ratio
(6000/300) times
0.5 $\mu\textrm{m}$
yielding
10 $\mu\textrm{m}$,
which does matter here, but,
rather, one encounters the maximum at a different location, namely
0.63 $\mu\textrm{m}$,
since obviously (6000/250) times
0.63 $\mu\textrm{m}$
yields
15.12 $\mu\textrm{m}$.
In the diagrams in Fig.\ 10, p.\ 293, one may recognize Wien's law,
though with some effort.
\item[(7)]
The multiplication of the Planck function with the wave length,
Georg Hoffmann's aggressively exaggerated issue, magnifies the
values at larger wave lengths (cf.\ Fig.\ 12, p.\ 295) but does not
suffice to yield equal areas under the curves. One always needs a
rescaling with factor 3.5 (in case of 5780 K, 290 K) and with factor
7.1 (in case of 6000 K, 250 K), respectively. Hiding this rescaling
intentionally in a scientific publication is definitely misleading
and may be characterized as scientific misconduct.
\end{itemize}
The authors emphasize: The multiplication of the Planck function
with the wave length - Georg Hoffmann's aggressively exaggerated
issue - was performed by the authors in their falsification paper
(cf.\ Fig.\ 12 and 13 of
Refs.~\refcite{GT2009,GT2007},
left hand
side and the text below)
\cite{GT2009,GT2007}.
\newpage
\subsection{\lq Twin Peaks\rq\ gallery}

\subsubsection{Overview}
In what follows, we try to shed some light on the background of the
\lq twin peaks\rq. Two questions are relevant:
\begin{enumerate}
\item
Why did Bakan and Raschke include diagrams in their paper
\cite{BR}
\begin{itemize}
\item
that are normally recognized as clearly wrong
both by meteorologists and by layman;
\item
that suggest the incorrect statement
the radiation intensity of the ground being equal
to the intensity  of the incoming solar radiation?
\end{itemize}
\item
Why did Bakan and Raschke not take Fig.\ 5.8 of the textbook by
Thomas and Stamnes
\cite{TS1999}
cited by themselves?
\end{enumerate}
Surely, Joshua Halpern knows the correct answer. Let us now evaluate
the diagrams from the perspective of a physicist.
\newpage

\subsubsection{Luther and Ellingson (1985)}
\begin{figure}[h]
\centerline{\includegraphics[scale=1]{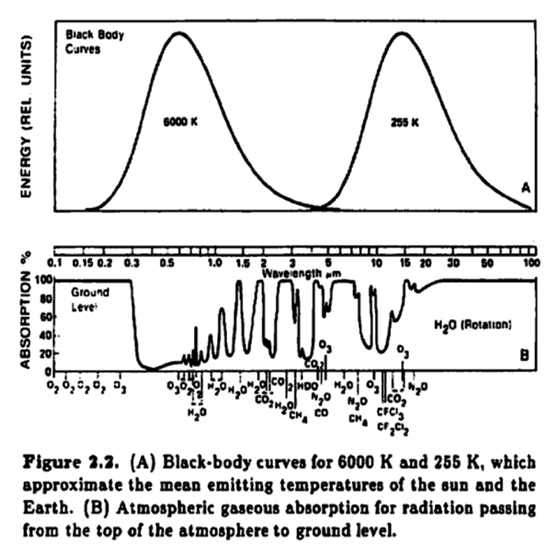}}
\vspace*{0pt} \caption{Diagram by Luther and Ellingson 1985, page
29} \label{Label1}
\end{figure}

The diagram (Fig.\ \ref{Label1}) in Luther and Ellingson (1985)
depicted on page 29 is incorrect
\cite{LE1985}.
To the ordinate
\lq\lq ENERGY (REL.\ UNITS)\rq\rq\ no units are attached. The text
in the diagram reads \lq\lq Black Body Curves\rq\rq, \lq\lq 6000
K\rq\rq, \lq\lq 255 K\rq\rq. The figure caption reads \lq\lq
blackbody emission at 6000 K ... and at 255\rq\rq.
\vspace*{20mm}%
\pagebreak\noindent

\subsubsection{Goody and Yung (1989)}
\begin{figure}[h]
\centerline{\includegraphics[scale=0.90]{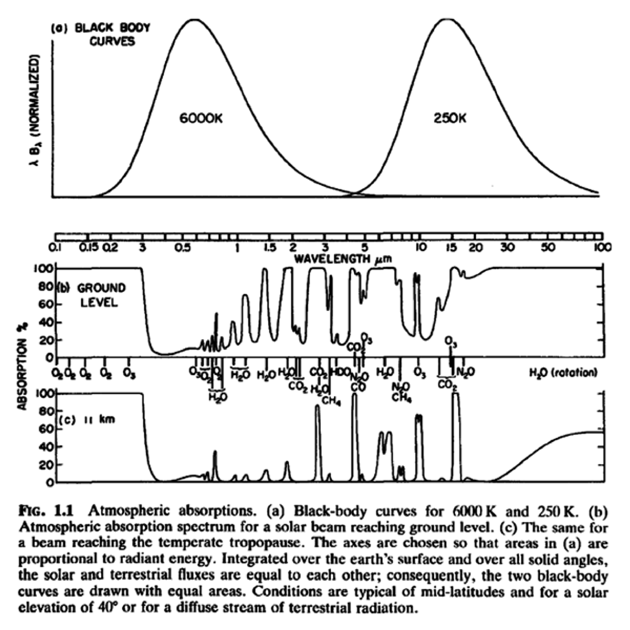}}
\vspace*{8pt} \caption{Diagram by Goody and Yung (1989), page 4}
\label{Label2}
\end{figure}

The diagram
(Fig.\ \ref{Label2})
in Goody and Yung (1989)
depicted on page 4 is incorrect
\cite{GY1989}.
This diagram looks similar to the
discussed diagram by Bakan and Raschke
depicted below.
To the ordinate
\lq\lq $\lambda \cdot B_\lambda$ (NORMALIZED)\rq\rq\
no units are attached.
The text in the diagram reads
\lq\lq BLACK BODY CURVES\rq\rq,
\lq\lq 6000 K\rq\rq,
\lq\lq 250 K\rq\rq.
The figure caption reads
\lq\lq Blackbody emission for 6000 K and 250 K\rq\rq.
\newpage%
\subsubsection{Thomas and Stamnes (1999)- Diagram 1}
\begin{figure}[h]
\centerline{\includegraphics[scale=1]{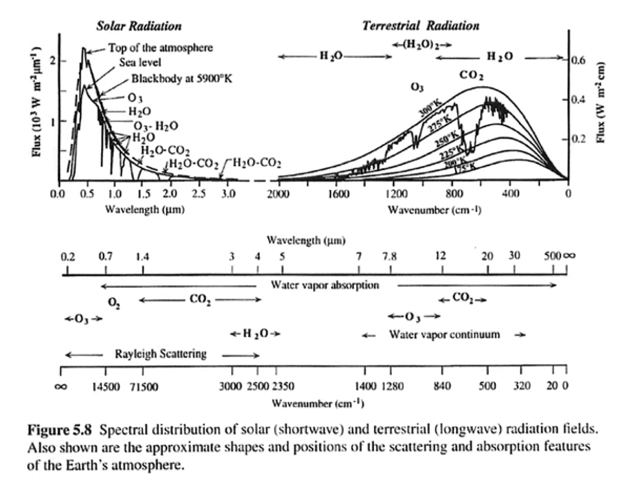}}
\caption{Diagram by Thomas and Stamnes (1999), page 149}
\label{Label3}
\end{figure}

The diagram (Fig.\ \ref{Label3}) in Thomas and Stamnes (1999)
depicted on page 149 is correct
\cite{TS1999}.
There are two
distinguished ordinate axes with two well-distinguished scales: The
scientific method at work.
\newpage%
\subsubsection{Thomas and Stamnes (1999) - Diagram 2}
\begin{figure}[h]
\centerline{\includegraphics[scale=0.70]{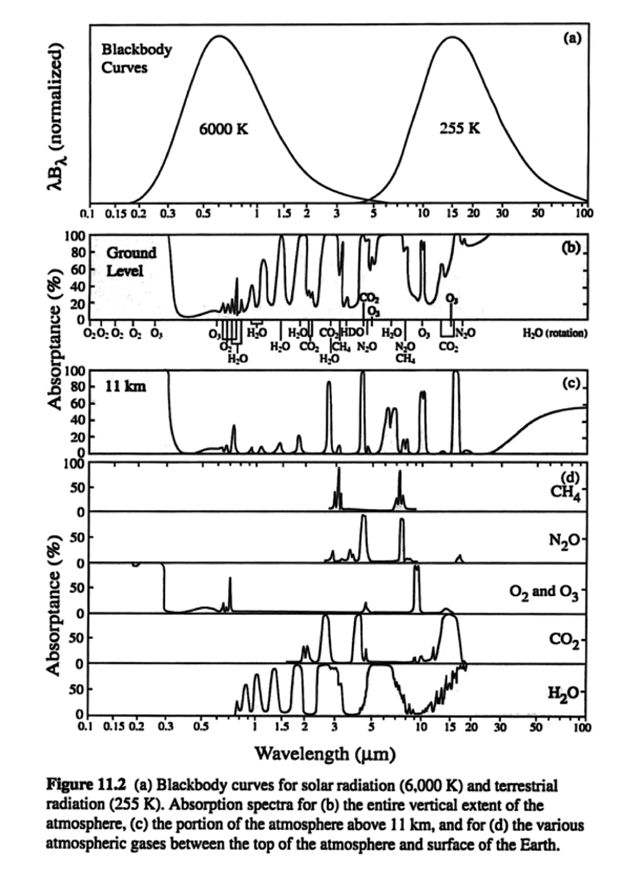}}
\caption{Diagram by Thomas and Stamnes (1999), page 420}
\label{Label4}
\end{figure}

Unfortunately,
on page 420 of their textbook
Thomas and Stamnes (1999)
include a diagram
(Fig.\ \ref{Label4})
that is incorrect
\cite{TS1999}.
It is reminiscent of that
depicted by Goody and Young (1989)
on page 4
\cite{GY1989}.
\newpage
\subsubsection{Bakan and Raschke (2002)}
\begin{figure}[h]
\centerline{\includegraphics[scale=1]{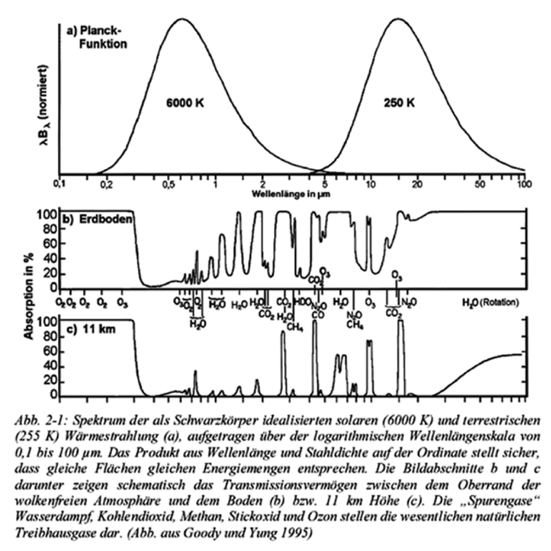}}
\caption{Diagram by Bakan and Raschke (2002), page 2} \label{Label5}
\end{figure}

Contrary to what Joshua Halpern et al.\
and Georg Hoffmann say, the
diagram (Fig.\ \ref{Label5}) depicted by Bakan and Raschke (2002) is
incorrectly scaled and incorrectly cited
\cite{BR}.
Although Bakan
and Raschke refer to the book by Goody and Yung (1980)
\cite{GY1989},
they introduce additional errors. For instance \lq\lq Planck
Function\rq\rq\ in this context is complete nonsense. On the bottom
line, they did not cite scientifically (correctly), that is, they
did not cite.
\newpage
\subsubsection{Petty (2006)}
\begin{figure}[h]
\centerline{\includegraphics[scale=1]{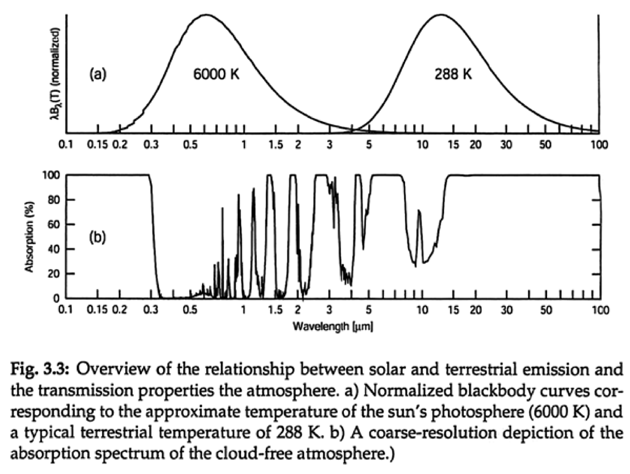}}
\caption{Diagram by Petty (2006), page 62} \label{Label6}
\end{figure}

The diagram (Fig.\ \ref{Label6}) in Petty (2006) depicted on page 65
is incorrect
\cite{Petty2006}.
To the ordinate \lq\lq $\lambda \cdot
B_\lambda$ (NORMALIZED)\rq\rq\ no units are attached. The figure
caption reads \lq\lq Normalized blackbody curves corresponding to
the approximate temperature of the sun's photosphere (6000 K) and a
typical terrestrial temperature of 288 K.\rq\rq.
\newpage
\subsubsection{Gerlich and Tscheuschner (2007, 2009)}
\begin{figure}[h]
\centerline{\includegraphics[scale=1]{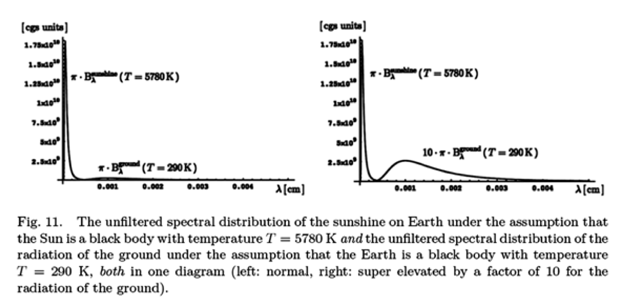}}
\caption{Diagram Fig.\ 12 by Gerlich and Tscheuschner (2007, 2009)}
\label{Label7}
\end{figure}

\begin{figure}[h]
\centerline{\includegraphics[scale=1]{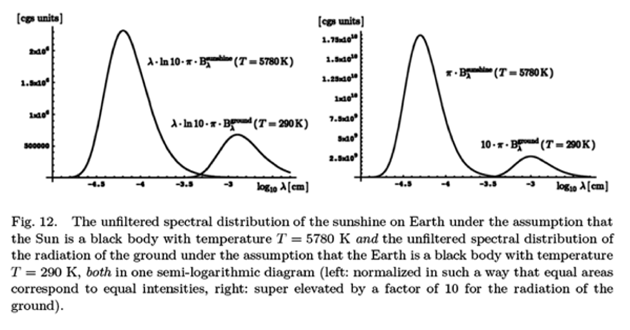}}
\caption{Diagram Fig.\ 12 by Gerlich and Tscheuschner (2007, 2009)}
\label{Label8}
\end{figure}

\begin{figure}[h]
\centerline{\includegraphics[scale=1]{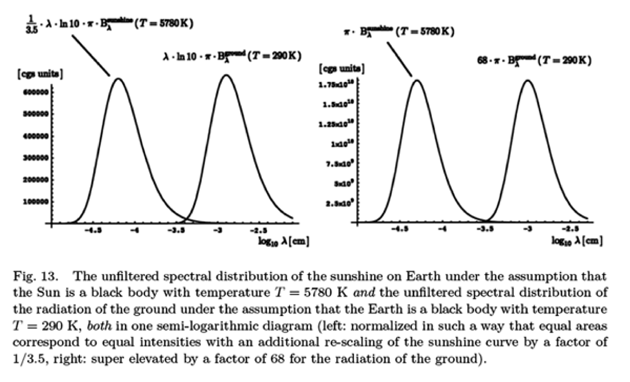}}
\caption{Diagram Fig.\ 13 by Gerlich and Tscheuschner (2007, 2009)}
\label{Label9}
\end{figure}

In our falsification paper
\cite{GT2009,GT2007}
we offered three
correct diagrams (Figures \ref{Label7}, \ref{Label8}, and
\ref{Label9}). They clearly show that the radiation from the ground
is much smaller than commonly suggested.
\newpage
\subsubsection*{4.2.8 (cont'd)\ \ Gerlich and Tscheuschner (2007, 2009)}
\newpage
\subsubsection*{4.2.8 (cont'd)\ \ Gerlich and Tscheuschner (2007, 2009)}
\newpage
\section{The Barometric Formulas}
\label{lapse}
\subsection{Overview}
In the speculative discussion around the existence of an atmospheric
natural greenhouse effect
\cite{NYT}
or the existence of an
atmospheric CO$_2$ greenhouse effect it is sometimes stated that the
greenhouse effect could modify the temperature profile of the
Earth's atmosphere. This conjecture is related to another popular
but incorrect idea communicated by some proponents of the global
warming hypothesis, namely that the temperatures of the
Venus are due to a greenhouse effect. For instance, in their book
\lq\lq Der Klimawandel. Diagnose, Prognose, Therapie\rq\rq\ (Climate
Change. Diagnosis, Prognosis, Therapy) \lq\lq two leading
international experts\rq\rq, Hans-Joachim Schellnhuber and Stefan
Rahmstorf, present a \lq\lq compact and under\-stand\-able
review\rq\rq\ of \lq\lq climate change\rq\rq\ to the general
public
\cite{RahmstorfSchellnhuber2006}.
On page 32 they explicitly
refer to the \lq\lq power\rq\rq\ of the \lq\lq greenhouse
effect\rq\rq\ on the Venus.

In
Refs.~\refcite{GerlichBaro,GT2010}
we explicitly derive the
approximate pressure profiles, density profiles, and temperature
profiles of an atmosphere, also called {\it barometric formulas\/}.
Our variant of a derivation goes beyond the common standard exercise
of a thermodynamics lecture, where commonly the discussion of the
underlying physical assumptions is missed.

These are:
\begin{quote}
\begin{enumerate}
\item The neglection of the electromagnetic field terms.
\item The independency of the wind velocities $\textbf{v}(\textbf{r},t)$
      on the location $\textbf{r}$.
\item The vanishing of the external force densities
      $\textbf{F}_{\mbox{\scriptsize\rm ext}}$.
\item The verticality of acceleration due to gravity.
\item The horizontality of the wind velocities.
\item The validity of ideal gas laws for the air of the atmosphere.
\item An isothermal atmosphere resp. an adiabatic atmosphere.
\item The independency of the specific heats of the gases on
      the absolute temperature in the case of an adiabatic
      atmosphere.
\end{enumerate}
\end{quote}
We depart from the Navier-Stokes equation and explicitly point our
attention on the physical assumptions disregarded elsewhere. By the
way, this derivation is a good example on how to apply the
magnetohydrodynamic equations regarded as redundant by some of our
critics. Furthermore, it explicitly shows that in physics an
application of formulas is valid {\it only in a finite space-time
region\/}. In addition, we show that the usual assumptions can be
relaxed leading to generalized formulas that hold even in the case
of horizontal winds.

A brief historical review of the barometric formula is given in
Ref.~\refcite{Berberan1996}.
The reader is also referred to the
textbook by Riegel and Bridger on \lq\lq Fundamentals of Atmospheric
Dynamics and Thermodynamics\rq\rq\ where the standard derivation of
the barometric formulas can be found
\cite{Riegel1992}.

\subsection{Results}
By combining hydrodynamics, thermodynamics, and imposing the above
listed assumptions for a planetary atmospheres one can compute the
temperature profiles of idealized atmospheres. In case of the
adiabatic atmosphere the decrease of the temperature with height is
described by a linear function with slope $-g/C_p$, where $C_p$
depends weakly on the molecular mass. As elucidated in our
paper
\cite{GT2009,GT2007}
mixtures of gases are covered in the
context of Gibbs thermodynamics. Since the measurable thermodynamic
quantities of a voluminous medium, in particular the specific heat
and the thermodynamic transport coefficients, naturally include the
contribution from radiative interactions, we cannot expect that a
change of concentration of a trace gas has any measurable effect. At
this point, it is important to remember that the barometric formulas
do not hold globally but have only a limited range of validity.

Let us return to the claim of Rahmstorf and Schellhuber that the
high venusian surface temperatures somewhere between 400 and 500
Celsius degrees are due to an atmospheric CO$_2$ greenhouse effect.
Of course, they are not. On the one hand, since the venusian
atmosphere is opaque to visible light, the central assumption of the
greenhouse hypothesis is not obeyed. On the other hand, if one
compares the temperature and pressure profiles of Venus and Earth,
one immediately sees, that they are both very similar. An important
difference is the atmospheric pressure on the ground, which is
approximately two orders higher than on the Earth. At 50 km altitude
the venusian atmospheric pressure corresponds to the normal pressure
on the Earth with temperatures at approximately 37 Celsius degrees.
However, things are extremely complex (volcanic activities, clouds
of sulfuric acid), such that we do not go in details
here
\cite{venus}.
\newpage
\section{Concluding Remarks}
In our falsification paper we have shown that the atmospheric CO$_2$
greenhouse effects as taken-for-granted concepts in global
climatology do not fit into the scientific realm of theoretical and
applied physics.

Halpern {\it et al.}
did not refute our conclusions. Rather, they
make false statements about the contents of our paper, on which they
erect their system of objections. Their main mistakes are:
\begin{enumerate}
\item
Halpern {\it et al.\/}
make false statements about the contents and the
rationale of our paper.
\item
Halpern {\it et al.\/}
do not understand what a physical effect really
is.
\item
Halpern {\it et al.\/}
- adapting Georg Hoffmann's view - apparently
do not know how to apply the second law of thermodynamics.
\item
Halpern {\it et al.\/}
do not understand our critique on the abuse of
diagrams in the context of simplistic radiative balance models.
\item
Halpern {\it et al.\/}
like many others do not understand that any
supposed warming effect (or cooling effect) cannot be derived from
spectroscopic analyses or radiative transfer equations.
\item
Halpern {\it et al.\/}
neither define a greenhouse effect nor offer a
mechanism how the concentration change of the trace gas CO$_{2}$
influences the climates.
\item
Halpern {\it et al.\/}
do not recognize the fundamental errors of the
paper by Bakan and Raschke.
\end{enumerate}
In summary, the paper of Halpern, Colose, Ho-Stuart, Shore, Smith,
and Zimmermann is unfounded
\cite{Halpern2010}.
\section*{Note Added in Proof}
\addcontentsline{toc}{section}{Note added in proof}

As Gerhard Kramm informed us recently, a correct version of the "twin peak"
diagrams can already be found on pape 17 in a forty years old textbook
on meteorology by Heinz Fortak \cite{Fortak1971}. We are very grateful for his hint.
\newpage
\section*{Acknowledgement}
\addcontentsline{toc}{section}{Acknowledgement}
We are grateful to all for giving us the opportunity to make our
point. Discussions with Dipl.-Met.\ Dr.\ Wolfgang Th\"{u}ne, Dipl-Ing.\
Paul Bossert, Prof.\ Dr.\ Gerhard Kramm (University of Alaska, Fairbanks), and
Klaus Ermecke (KE Research) are acknowledged. Gerhard Gerlich thanks
Grant Petty sending him an electronic copy of his
book
\cite{Petty2006}.
Ralf D. Tscheuschner acknowledges an
interesting discussion (10. January 2008) with Prof.\ Dr.\
Friedrich-Wilhelm Gerstengarbe (PIK), Dr.\ \"Osterle (PIK), Prof.\ Dr.\
Peter C.\ Werner (PIK), and Dipl.-Ing.\ Michael Limburg (EIKE) on
evaluation of terrestrial temperature measurements. Thanks to
Dipl.-Phys.\ Dr.\ Manfred Dinter for a critical last-minute reading.

We thank the referee for his constructive suggestions.
\newpage

\end{document}